\documentclass[conference]{IEEEtran}
\IEEEoverridecommandlockouts
\usepackage{cite}
\usepackage{amsmath,amssymb,amsfonts}
\usepackage{algorithmic}
\usepackage{graphicx}
\usepackage{textcomp}
\usepackage{xcolor}
\usepackage{amsmath,amsfonts}
\usepackage{algorithmic}
\usepackage{algorithm}
\usepackage{array}
\usepackage[caption=false,font=normalsize,labelfont=sf,textfont=sf]{subfig}
\usepackage{textcomp}
\usepackage{stfloats}
\usepackage{url}
\usepackage{verbatim}
\usepackage{graphicx}
\usepackage{cite}
\usepackage{booktabs}
\usepackage{tgpagella}

\usepackage{amssymb}
\usepackage{amsmath}
\usepackage{subfig}
\usepackage{hyperref}
\hypersetup{colorlinks=true,citecolor=blue}
\usepackage{graphicx}
\usepackage{mathrsfs}
\usepackage{bbding}
\usepackage{rotating}
\usepackage{multirow}
\usepackage{marginnote}
\usepackage{overpic}
\usepackage{colortbl}
\usepackage{pifont}
\usepackage{booktabs}
\usepackage[labelfont=bf]
{caption}
\usepackage{bbding}

\def\ourmodel{FocusNet}
\definecolor{green}{HTML}
{1D7A00}

\def\BibTeX{{\rm B\kern-.05em{\sc i\kern-.025em b}\kern-.08em
    T\kern-.1667em\lower.7ex\hbox{E}\kern-.125emX}}
\begin{document}

\title{FocusNet: Transformer-enhanced Polyp Segmentation with Local and Pooling Attention 


}

\author{
\IEEEauthorblockN{
Jun Zeng\IEEEauthorrefmark{1},
KC Santosh\IEEEauthorrefmark{2},
Deepak Rajan Nayak\IEEEauthorrefmark{3},
Thomas de Lange\IEEEauthorrefmark{4},
Jonas Varkey\IEEEauthorrefmark{4},
Tyler Berzin\IEEEauthorrefmark{4},
Debesh Jha\IEEEauthorrefmark{2}}
\IEEEauthorblockA{\IEEEauthorrefmark{1}Chongqing University of Posts and Telecommunications, Chongqing, China}
\IEEEauthorblockA{\IEEEauthorrefmark{2}University of South Dakota, Vermillion, SD, USA}
\IEEEauthorblockA{\IEEEauthorrefmark{3}Malaviya National Institute of Technology Jaipur, Rajasthan, India}
\IEEEauthorblockA{\IEEEauthorrefmark{4}Harvard Medical School, Boston, MA, USA}
\IEEEauthorblockA{\IEEEauthorrefmark{5}University of Gothenburg, Gothenburg, Sweden}
}

\maketitle

\begin{abstract}
Colonoscopy is vital in the early diagnosis of colorectal polyps. Regular screenings can effectively prevent benign polyps from progressing to CRC. While deep learning has made impressive strides in polyp segmentation, most existing models are trained on single-modality and single-center data, making them less effective in real-world clinical environments. To overcome these limitations, we propose FocusNet, a Transformer-enhanced focus attention network designed to improve polyp segmentation. FocusNet incorporates three essential modules: the Cross-semantic Interaction Decoder Module (CIDM) for generating coarse segmentation maps, the Detail Enhancement Module (DEM) for refining shallow features, and the Focus Attention Module (FAM), to balance local detail and global context through local and pooling attention mechanisms. We evaluate our model on PolypDB, a newly introduced dataset with multi-modality and multi-center data for building more reliable segmentation methods. Extensive experiments showed that FocusNet consistently outperforms existing state-of-the-art approaches with a high dice coefficients of 82.47\% on the BLI modality, 88.46\% on FICE, 92.04\% on LCI,  82.09\% on the NBI and 93.42\% on WLI modality, demonstrating its accuracy and robustness across five different modalities. The source code for FocusNet is available at \url{https://github.com/JunZengz/FocusNet}.

\end{abstract}

\begin{IEEEkeywords}
Colonoscopy, Polyp Segmentation, Transformer, Attention Mechanism.
\end{IEEEkeywords}

\section{Introduction}
Colorectal cancer (CRC) is the third most common cancer globally and the second leading cause of cancer-related deaths in both men and women~\cite{siegel2023colorectal}. In the U.S., CRC is the second leading cause of cancer deaths, with an estimated 53,010 deaths in 2024, a 460-death increase from 2023. CRC is now the leading cause of death in men under 50 and the second in women~\cite{siegel2024cancer}. The incidence in younger individuals is rising by 1-2\% annually, with mortality increasing by 1\%, largely due to delayed detection. 

CRC typically begins from benign polyps in the colorectal mucosa, with adenomatous polyps making up about 70\% of all colorectal polyps and recognized as precursors to CRC. Not all benign polyps progress to CRC, and polyps smaller than 10 mm are often overlooked by physicians~\cite{hixson1990prospective}. Intestinal folds and poor visibility can also hinder detection. While colonoscopy is effective, small polyps can be missed due to operator dependence. Computer-assisted segmentation methods could reduce missed detections, making accurate segmentation, especially for small or camouflaged polyps, crucial for improving diagnostic accuracy.

The advancement of deep learning technologies has significantly accelerated research in polyp segmentation. The success of U-Net~\cite{ronneberger2015u} in medical image segmentation has established the encoder-decoder architecture as a dominant framework, leading to the development of a variety of enhanced methods, including ResUNet~\cite{zhang2017resunet}, ResUNet++~\cite{jha2019resunet++}, DoubleU-Net~\cite{jha2020doubleu}, TransU-Net~\cite{chen2021transunet}, and TransResU-Net~\cite{tomar2022transresu}. The integration of attention mechanisms~\cite{tomar2022tganet,nguyen2021ccbanet, kim2021uacanet, duc2022colonformer, yin2024rsaformer} has further refined boundary precision in polyp segmentation. Other methods~\cite{srivastava2021msrf, srivastava2022gmsrf, tomar2022automatic, liu2024multi} have also demonstrated significant improvements by leveraging multi-scale feature aggregation. Recently, transformer-based backbones, such as Swin Transformer~\cite{liu2021swin} and PVTv2~\cite{wang2021pvtv2}, have achieved significant success in enhancing the generalization capabilities of polyp segmentation models. Notably, Polyp-PVT~\cite{dong2021polyp} has shown that the PVTv2 backbone is highly effective for feature extraction in polyp segmentation. 

Clinical practice often relies on multiple endoscopic imaging techniques to enhance visibility and diagnostic accuracy under different conditions. However,  most of the algorithms found in the literature only developed using White light imaging (WLI), and there is a critical gap in their applicability across other clinically relevant imaging modalities. Advanced imaging techniques such as Blue Laser Imaging (BLI), Flexible Spectral Imaging Color Enhancement (FICE), Linked Color Imaging (LCI), and Narrow Band Imaging (NBI) are frequently used along with the WLI during colonoscopy procedures for improving polyp detection, especially for flat, small hyperplastic, sessile serrated polyp or camouflaged lesions. Each modality emphasizes different visual characteristics of the mucosa—such as vascular patterns or surface textures—necessitating models that can adapt to diverse visual domains.



A segmentation model that performs reliably well across all five modalities can be crucial for real-world deployment in real-world clinical environments. To address this need, we propose \ourmodel, a novel focus attention network that combines semantic interaction, detail enhancement, and multi-scale attention mechanisms for accurate and generalizable polyp segmentation. To our knowledge, this is the first approach to evaluate polyp segmentation across multiple modalities and multiple centers. We validate the effectiveness of our approach and SOTA methods on the PolypDB dataset and evaluate the performance in both modality-wise and center-wise settings. Our contributions can be summarized as follows:

\begin{enumerate}
\item 
We propose \textit{FocusNet}, a novel network architecture aimed to achieve precise polyp segmentation in multi-modalities, which combines local and pooling attention to capture detailed information from shallow features.

\item 
We introduce three modules: {\em Cross-semantic Interaction Decoder Module} (CIDM), {\em Detail Enhancement Module} (DEM), and {\em Focus Attention Module} (FAM), all aimed at improving segmentation performance. The CIDM module enables the interaction and fusion of multi-level deep features, while the DEM module enhances fine-grained details by processing shallow features through multi-path deformable convolutions. Additionally, the FAM module is designed to balance the extraction of local details with global context semantic understanding, resulting in robust polyp feature representations.

\item We conduct extensive evaluations on PolypDB using both modality-wise and center-wise settings, comparing CNN-based, Transformer-based, and Mamba-based methods. FocusNet consistently achieves SOTA performance in both qualitative and quantitative analyses, highlighting its strong learning and generalization capabilities.
\end{enumerate}

\section{Method}\label{sec:Methods} 
\begin{figure*}[!t]
	\centering
	\includegraphics[width=0.95\linewidth]{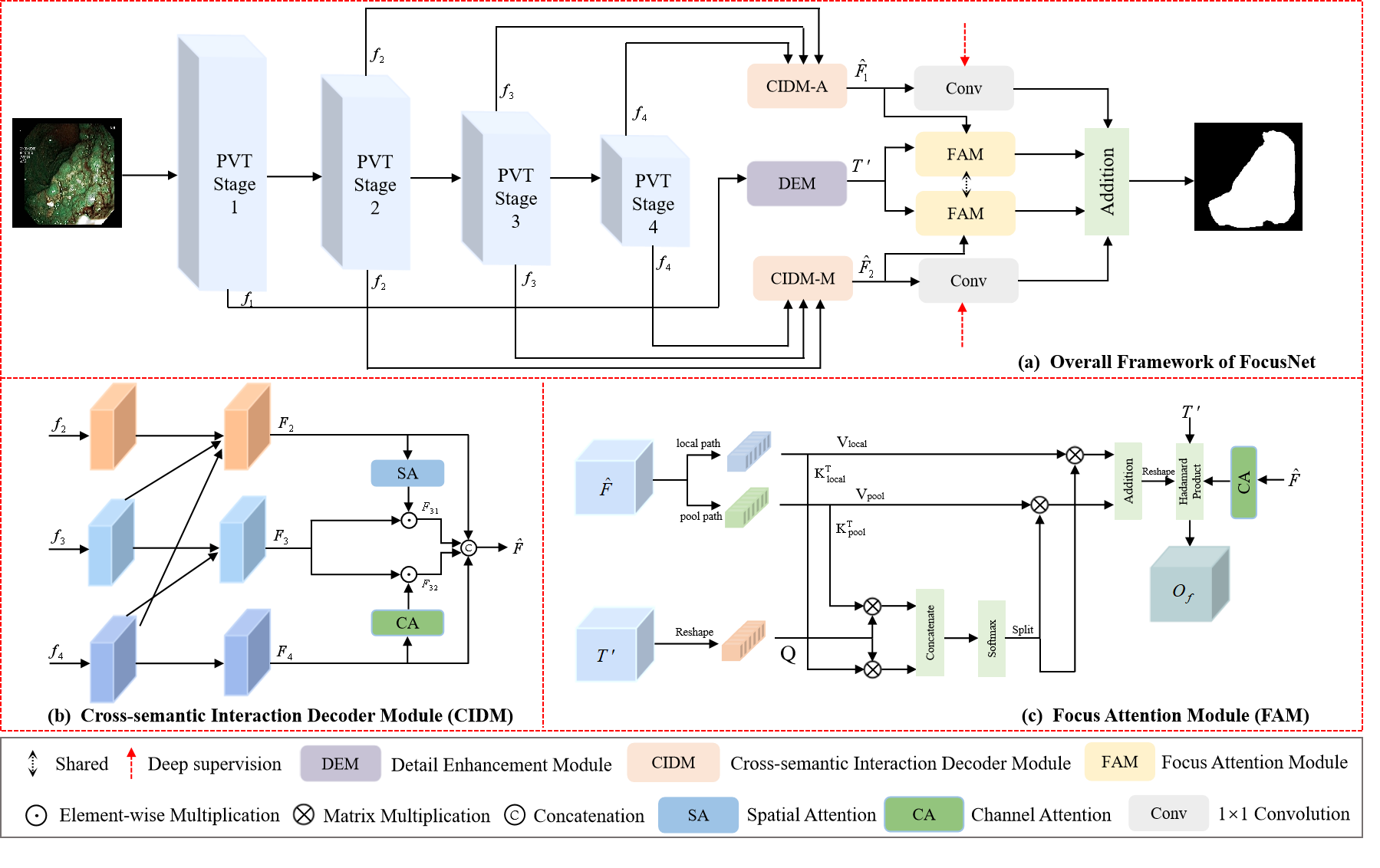}
	\caption{ Overview of the proposed FocusNet architecture. The network comprises four main components: (1) a PVTv2~\cite{wang2022pvt} backbone that extracts multi-scale hierarchical features from the input image; (2) a Detail Enhancement Module (DEM) that refines shallow features ($f_1$) to preserve fine-grained boundary information; (3) two Cross-semantic Interaction Decoder Modules (CIDM-M and CIDM-A) that process deep features ($f_2$, $f_3$, $f_4$) to generate coarse segmentation maps by enhancing semantic fusion across multiple levels; and (4) a Focus Attention Module (FAM) that combines outputs from DEM and CIDMs using both local and pooling attention to capture detailed spatial features and global context. The FAM produces segmentation maps ($P_3$, $P_4$), which are fused with coarse predictions ($P_1$, $P_2$) to generate the final segmentation output ($\Hat{P}$) for accurate and robust polyp segmentation.}
	\label{fig:framework}
\end{figure*}

\subsection{Overview }\label{sec:overview}
The network architecture of \ourmodel~is shown in Figure~\ref{fig:framework}.  FocusNet consists of four main components: a pyramid vision transformer (PVTv2)~\cite{wang2022pvt} backbone, a Detailed enhancement module (DEM), two cross semantic interaction decoder modules(CIDM-A and CIDM-M), and a focus attention (FAM) module. The PVTv2 is used as the backbone to extract multi-scale features $f_i$ with a resolution of $\frac{H}{2^{i+1}} \times \frac{W}{2^{i+1}} \times C_{i}$ from the input polyp image, where $C_{i} \in \left\{64, 128, 320, 512 \right\}$ and $i \in \left\{1, 2, 3, 4 \right\}$. The DEM module processes the shallow features $f_1$ from the first transformer block of PVTv2 and generates a fine-grained feature map $T'$. The CIDM-M module takes the deep features $\left\{f_2, f_3, f_4\right\}$ as input and generates the intermediate feature map $\hat{F_1}$ and the segmentation map $P_1$. Similarly, $\left\{f_2, f_3, f_4\right\}$ are passed into the CIDM-A module, producing $\hat{F_2}$ and $P_2$. Subsequently, the feature maps $\left\{ \hat{F_1}, \hat{F_2}\right\}$  and $T'$ are fed into the FAM module, which generates two additional segmentation maps $P_3$ and $P_4$. Finally, all the segmentation maps $\left\{P_1, P_2, P_3, P_4 \right\}$ are fused through an \textbf{element-wise} addition operation to obtain the final prediction map $\Hat{P}$.

\subsection{Cross-semantic Interaction Decoder Module }\label{sec:CIDM}
The hierarchical features generated by the encoder are critical to polyp segmentation. These features contain abstract information that plays a crucial role in polyp localization. To bridge the gap between hierarchical features and foster their integration, we propose a CIDM. In CIDM, we utilize the later three levels of the encoder $ \left\{f_{2}, f_{3}, f_{4} \right\}$ to generate the coarse polyp segmentation map.

The architecture of CIDM is illustrated in Figure  \ref{fig:framework}. Firstly, the feature maps $f_2$, $f_3$, and $f_4$ are transformed into the same number of channels using 1x1 convolutions, respectively. Next, $f_4$ is upsampled by a factor of 2 and element-wise multiplied with $f_3$. After that, both $f_4$ and $f_3$ are upsampled by factors of 4 and 2, respectively, and element-wise multiplied with $f_2$. The process can be summarized as follows:



\begin{equation}  \label{eq:1}
	\left\{
	\begin{array}{rcl}
		F_4 &=& {\text{Up}_4}(f_4), \\
		F_3 &=& {\text{Up}_2}(\text{Conv}({\text{Up}_2}(f_4)) \odot f_3), \\
		F_2 &=& \text{Conv}({\text{Up}_4}({f_4}) \odot \text{Conv}({\text{Up}_2}(f_3)) \odot f_2,
	\end{array}
	\right.
\end{equation}
where ${\text{Up}_i}$ denotes an upsampling operation with a factor of ${i}$; Conv represents a convolution operation; $\odot$ denotes element-wise multiplication. After that, $F_4$ is first processed by channel attention (CA)~\cite{woo2018cbam}, then multiplied element-wise with $F_3$ to generate $F_{31}$. Similarly, $F_2$ is processed by spatial attention (SA)~\cite{hu2018squeeze} and multiplied element-wise with $F_3$ to produce $F_{32}$. Finally, $F_2$, $F_{31}$, $F_{32}$, and $F_4$ are concatenated and passed through a convolution layer to generate the final feature map $\hat{F}$ and the coarse segmentation map $P$. $Concat(\cdot)$ is a concatenation
operation along the channel dimension. This process can be summarized as follows:
\begin{equation}  \label{eq:2}
	\left\{
	\begin{array}{lcl}
		F_{31} &=& F_3 \odot CA(F_4) \\
		F_{32} &=& F_3 \odot SA(F_2) \\
		\hat{F} &=& \text{Concat}(F_2, F_{31}, F_{32}, F_4),
	\end{array}
	\right.
\end{equation}
where $CA(\cdot)$ stands for channel attention, $SA(\cdot)$ denotes spatial attention. The process described above outlines the details of CIDM-M. In addition, we modify Equation \eqref{eq:1} by replacing all element-wise multiplication operations with addition operations. This modification results in a new variant of this module, referred to as CIDM-A.


\subsection{Detail Enhancement Module}\label{sec:DEM}
Low-level features contain rich detail information, such as colors, contours, and the appearance of polyps. Highlighting these details is beneficial for subsequent processing, as it helps to complement edge information. To achieve this goal, we propose a detail enhancement module (DEM), which is used to refine low-level features and enhance the fine-grained details. 




First, we partition the first-stage feature output $f_1$ from the encoder into four equal parts along the channel dimension. Each part is then fed into a separate branch, where each branch consists of a sequence of 1$\times$3 and 3$\times$1 deformable convolutions. These convolutions are followed by Batch Normalization (BN) and ReLU activation. The outputs of the four branches are concatenated along the channel dimension to capture diverse directional spatial features. 
The process can be expressed as follows:
\begin{equation}  \label{eq:3}
    T_i = BR(DConv_{1 \times 3}(BR(DConv_{3 \times 1}(t_i)))) 
\end{equation}

\begin{equation}  \label{eq:4}
    T = Concat(T'_1, \dots, T_k), \quad k = 4
\end{equation}
where $DConv_{1 \times 3}$ and $DConv_{3 \times 1}$ refer to deformable convolutional layers with kernels of size 1$\times$3 and 3$\times$1, respectively. BR refers to BN followed by ReLU activation.
Then the concatenated feature map is passed through an efficient channel attention (ECA)~\cite{wang2020eca} to refine the channel-wise dependencies, ensuring that important features are effectively highlighted. Finally, the enhanced feature map is processed through a 1×1 convolutional layer, followed by BN and ReLU activation to produce the final output of the module. 
The above process can be formulated as follows:
\begin{equation}  \label{eq:5}
    T' = BRU(ECA(Conv_{1\times1}(T)))
\end{equation}
where $ECA(\cdot)$ refers to the Eficient Channel Attention, $BRU(\cdot)$ stands for a 1$\times1$ convolution, followed by BN and ReLU activation, and $Conv_{1 \times 1}$ refers to a convolutional layer with a kernel of size 1$\times$1.

\subsection{Focus Attention Module}\label{sec:FAM}
Human visual perception has higher acuity for features around the focal point but lower attention to distant features. The combination of fine-grained attention and coarse-grained attention can align with biological vision~\cite{shi2024transnext}. In this paper, 
we propose a focus attention module to balance the local details and global context by combining local and pooling attention for precise polyp segmentation.




The structure of the FAM is shown in Figure \ref{fig:framework}.
The feature map $T'$ is transformed into 32 channels using a 1x1 convolutional layer and then fed into FAM as input. In FAM, a dual-path design is employed, which consists of a local path and a pool path. A linear layer is applied to $T'$ to generate a shared query $Q$ for both paths. In each path, the feature map $\hat{F}$ passes through two separate linear layers to produce independent key and value representations. Next, we compute the similarity scores, $S_{local}$ and $S_{pool}$, by performing matrix multiplication between the query and the key in both paths. This process can be formulated as follows:
\begin{equation}  \label{eq:6}
	\left\{
	\begin{array}{lcl}
		K_{local}, V_{local} &=& \gamma(\alpha(Linear(\hat{F})))
         \\
		K_{pool}, V_{pool} &=& \gamma(Linear(\delta(\hat{F})))
	\end{array}
	\right.
\end{equation}
where $\gamma(\cdot)$ represents the operation that splits the input tensor into two parts along the channel dimension. $\alpha(\cdot)$ denotes the reshape operation that generates local tokens based on the local window size. $\delta(\cdot)$ denotes the adaptive average pooling operation that generates pooling tokens based on the pooling window size. Here, we set both the local and pooling window size to 3. These similarity scores are then concatenated and passed through a softmax function to obtain the attention weights, which can be expressed as:
\begin{equation} \label{eq:7}
	Attn = Softmax(Concat(QK^T_{local}, QK^T_{pool}))
\end{equation}

We then split the attention weights along the channel dimension to obtain the local and pooling attention weights. Each attention weight is multiplied by its corresponding value, followed by an addition operation to produce the refined feature $O_{r}$. This process can be summarized as follows:
\begin{equation} \label{eq:8}
	O_r = \beta(Attn_{local}V_{local} + Attn_{pool}V_{pool})
\end{equation}
where $\beta(\cdot)$ denotes a linear layer followed by a dropout operation. Subsequently, the channel attention is applied to the input feature $\hat{F}$, yielding the refined feature $\hat{F}_{ca}$. The feature $F_{ca}$ is element-wise multiplied by the refined feature $O_{r}$ and the result is further element-wise multiplied by the input feature $T'$, generating the contextual feature $O_f$. Finally, $O_f$ is passed through convolutional layers and added to the coarse segmentation map $P$ to obtain the output segmentation map $P'$. This process can be formulated as:
\begin{equation} \label{eq:9}
	O_{f} = O_r \odot \hat{F}_{ca} \odot T'
\end{equation}

\begin{equation} \label{eq:10}
	P' = P + Conv_{1\times1}(Conv_{3\times3}(Conv_{3\times3}(O_f))
\end{equation}
where $\odot$ denotes hadamard product. Here, both $\hat{F_1}$ from CIDM-M and $\hat{F_2}$ from CIDM-A are fed into the shared FAM module to generate the corresponding segmentation maps. 



\section{Experiments}\label{sec:Experiments}

\subsection{Datasets and Evaluation Metrics}\label{sec:Datasets and Benchmarks}
\begin{figure}[!t]
	\centering
	\includegraphics[width=0.9\linewidth]{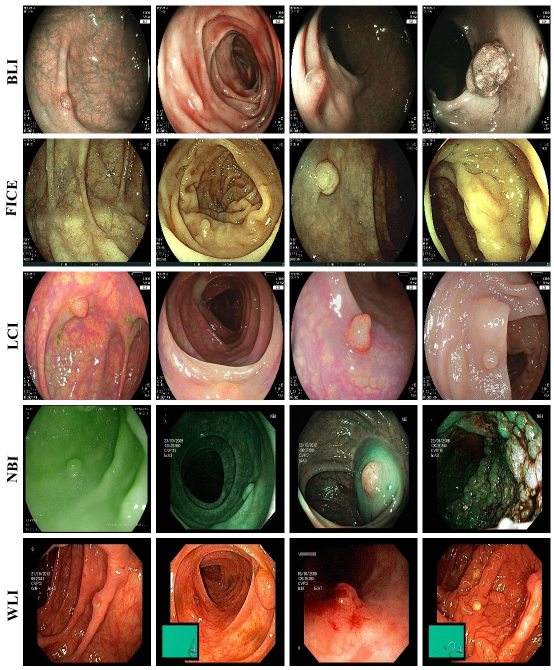}
	\caption{Samples from various modalities in the PolypDB dataset are shown in each row, with the corresponding modality descriptions provided on the left of each row.}
        \label{fig:SamplesfromDiffModalities}
\end{figure}

PolypDB~\cite{Jha2024PolypDB} is a large-scale, multi-center, multi-modality polyp dataset. The multi-center data is sourced from three medical centers in Norway, Sweden, and Vietnam. The multi-modality data includes image modalities from Blue Laser Imaging (BLI), Flexible Spectral Imaging Color Enhancement (FICE), Linked Color Imaging (LCI), Narrow Band Imaging (NBI), and White Light Imaging (WLI). The dataset comprises 3,934 polyp images and corresponding high-quality masks, which were annotated and verified by gastroenterology experts. Figure~\ref{fig:SamplesfromDiffModalities} showcases representative samples from different modalities. In the modality-wise setting, the 3588 WLI images were split into 2870 for the training set, 358 for the validation set, and 360 for one of the test datasets. Other modality images, including BLI (70), FICE (70), LCI (60), and NBI(146), were included as the remaining test datasets. In the center-wise setting, only the WLI images were used to ensure modality fairness for it's common in all three centers. The 2558 WLI images from the Simula center were divided into 2046 for the training set, 255 for the validation set, and 257 for one of the test datasets. Other center images, including BKAI (1000) and Karolinska (30), were included as the remaining test datasets.

To quantitatively evaluate the performance of various deep learning architectures, we employed seven evaluation metrics: Mean Intersection over Union (mIoU), Mean Dice Coefficient (mDSC), Recall, Precision, Accuracy, F2 Score, and P-values. While the first six metrics measure overlap-based and pixel-wise classification performance, we also include  P-values for statistical significance testing to confirm whether the observed performance differences between models are statistically meaningful. We also present computational complexity and inference speed to evaluate the efficiency of each method.








\subsection{Implementation Details}

We implemented FocusNet and all other SOTA using the PyTorch framework, with training conducted on a single NVIDIA Tesla V100 GPU. During training, we used a batch size of 16 and set the learning rate to $1\times10^{-4}$. To enhance the model's generalization, we applied robust data augmentation techniques, including random rotation, horizontal and vertical flipping, and coarse dropout. During the training phase, we train the network using a combination of binary cross-entropy loss and dice loss. In addition, the maximum number of epochs is set to 500 and an early stopping strategy with a patience value of 50 was employed to prevent overfitting.

\begin{table*}[t!]
	\centering
	\caption{Quantitative results of different methods on the PolypDB dataset with modality-wise setting. Here, \textcolor{red}{red} refers to the highest score, \textcolor{blue}{blue} represents to the second highest and \textcolor{green}{green} denotes to the third-highest score.}
    
    \label{tab:modality-wise}
	\begin{tabular}{clccccccc}
		\toprule
		
	Dataset	& Methods & mIoU ${\uparrow}$ & mDSC ${\uparrow}$  &  Recall ${\uparrow}$ & Precision ${\uparrow}$ & Accuracy ${\uparrow}$ & F2 ${\uparrow}$ & P-values
    ${\downarrow}$ \\
		\hline

    \multirow{8}{*}{
    \shortstack{PolypDB \\ (BLI)}} & 
    
    U-Net (MICCAI'15)
    ~\cite{ronneberger2015u}  &0.3591	&0.4143	&0.4508	&0.7998	&0.9664 &0.4212 &4.282e-13\\
    
    &U-Net++ (TMI'19)
    ~\cite{zhou2018unetplus} 	&0.3686	&0.4223	&0.4497	&0.8654	&0.9680 &0.4281 
    &6.723e-13\\

    &PraNet (MICCAI'20)
    ~\cite{fan2020pranet} &0.6582	&0.7353	&0.7660	&0.8430	&0.9779 &0.7342
    &9.162e-04\\

    
    &CaraNet (MIIP'22) 
    ~\cite{lou2022caranet}
    &0.6873	&0.7607	&0.7891	&\textcolor{red}{0.8769}&\textcolor{blue}{0.9845} &0.7645
    &1.321e-02\\
   
    &PVT-CASCADE (WACV'23)
    ~\cite{rahman2023medical}
    &\textcolor{blue}{0.7258}	&\textcolor{blue}{0.8017}	&\textcolor{blue}{0.8208}	&\textcolor{blue}{0.8684}	&\textcolor{green}{0.9838} 
    &\textcolor{blue}{0.8006}
    &1.574e-01\\
    

    &Mamba-UNet (arXiv'24)     ~\cite{wang2024mambaUnet}
    &0.6752  &0.7572  &0.7786 	&0.8303	&0.9808 &0.7554 &1.856e-03   \\

    &RMAMamba-T (arXiv'25)     ~\cite{zeng2025rmamamba}
    &\textcolor{green}{0.7038} &\textcolor{green}{0.7822}	&\textcolor{green}{0.8123}	&\textcolor{green}{0.8602}	&0.9834 &\textcolor{green}{0.7852} &1.046e-02\\

  
    &\textbf{\ourmodel~(Ours)} & \textcolor{red}{0.7460} & \textcolor{red}{0.8247} & \textcolor{red}{0.8472} & 0.8404   &\textcolor{red}{0.9887} &\textcolor{red}{0.8344} &-\\
    
    \hline

    \multirow{8}{*}{
    \shortstack{PolypDB \\ (FICE)}} &
    
    U-Net (MICCAI'15)
    ~\cite{ronneberger2015u}  &0.2526	&0.2773	&0.2701	&0.9167	&0.9553 &0.2713
    &8.887e-20\\
    
    &U-Net++ (TMI'19)
    ~\cite{zhou2018unetplus} 	&0.3491	&0.3830	&0.3687	&\textcolor{red}{0.9610}	&0.9574 &0.3727
    &6.103e-15\\
    
    &PraNet (MICCAI'20)
    ~\cite{fan2020pranet} &0.7182	&0.7780	&0.7728	&\textcolor{blue}{0.9249}	&0.9857 &0.7725 &4.542e-04\\

    
    &CaraNet (MIIP'22) 
    ~\cite{lou2022caranet}
    &0.7476	&0.8052	&0.8032	&0.9216 &0.9878	&0.8022
    &2.074e-03\\

    &PVT-CASCADE (WACV'23)
    ~\cite{rahman2023medical}
    &\textcolor{green}{0.8050}	&\textcolor{green}{0.8650}	&\textcolor{green}{0.8592}	&\textcolor{green}{0.9230}	&\textcolor{blue}{0.9910} &\textcolor{green}{0.8595}
    &2.038e-01\\


    &Mamba-UNet (arXiv'24)     ~\cite{wang2024mambaUnet}
    &0.7786  &0.8364  &0.8410 	&0.8753	&0.9874 &0.8366 &1.650e-02\\

    &RMAMamba-T (arXiv'25)     ~\cite{zeng2025rmamamba}
    &\textcolor{blue}{0.8144} &\textcolor{blue}{0.8766}	&\textcolor{red}{0.8872}	&0.8945	&\textcolor{green}{0.9904} &\textcolor{red}{0.8812} &6.467e-01\\

    
    &\textbf{\ourmodel~(Ours)} & \textcolor{red}{0.8253} & \textcolor{red}{0.8846} & \textcolor{blue}{0.8802} & 0.9085   &\textcolor{red}{0.9914} &\textcolor{blue}{0.8806} &-\\

    \hline

    \multirow{8}{*}{
    \shortstack{PolypDB \\ (LCI)}} 
    
    &U-Net (MICCAI'15)
    ~\cite{ronneberger2015u} &0.6948	&0.7677	&0.8393	&0.8299	&0.9851 &0.7947
    &9.115e-05\\
    
    &U-Net++ (TMI'19)
    ~\cite{zhou2018unetplus} 	&0.6421	&0.7183	&0.8072	&0.8017	&0.9749 &0.7528
    &5.526e-06\\
    
    &PraNet (MICCAI'20)
    ~\cite{fan2020pranet} &0.8313	&0.8935	&0.9107	&0.9150	&0.9930 &0.9015
    &1.164e-01\\

    
    &CaraNet (MIIP'22) 	
    ~\cite{lou2022caranet}
    &0.8357	&0.9013	&0.9054	&\textcolor{red}{0.9219}	&\textcolor{blue}{0.9936} &0.9021 &1.208e-01\\

    &PVT-CASCADE (WACV'23) 
    ~\cite{rahman2023medical}
    &0.8362	&0.8969	&0.9190	&0.9113	&0.9930 &0.9079
    &1.350e-01\\ 

    
    &Mamba-UNet (arXiv'24)     ~\cite{wang2024mambaUnet}
    &\textcolor{green}{0.8404}  &\textcolor{green}{0.9044}  &\textcolor{blue}{0.9379} 	&0.8951	&0.9908 &\textcolor{green}{0.9207} &8.935e-02\\

    &RMAMamba-T (arXiv'25)     ~\cite{zeng2025rmamamba}
    &\textcolor{blue}{0.8568} &\textcolor{blue}{0.9145}	&\textcolor{red}{0.9381}	&\textcolor{green}{0.9130}	&\textcolor{green}{0.9931} &\textcolor{blue}{0.9258} &5.366e-01\\

    
    &\textbf{\ourmodel~(Ours)} &\textcolor{red}{0.8613} 
    &\textcolor{red}{0.9204} 
    &\textcolor{green}{0.9378} 
    &\textcolor{blue}{0.9182}  
    &\textcolor{red}{0.9941} 
    &\textcolor{red}{0.9292}
    &-\\
        
    \hline
    
    \multirow{8}{*}{\shortstack{PolypDB \\ (NBI)}} 
    
    &U-Net (MICCAI'15)
    ~\cite{ronneberger2015u} &0.3269	&0.3968	&0.4130	&0.7724	&0.9250 &0.3992
    &1.233e-27\\
    
    &U-Net++ (TMI'19)
    ~\cite{zhou2018unetplus} 	&0.3154	&0.3774	&0.3970	&0.8367	&0.9210 &0.3789
    &2.816e-28\\
    
    &PraNet (MICCAI'20)
    ~\cite{fan2020pranet} &0.6360	&0.7131	&0.7332	&0.8419	&0.9590 &0.7164
    &1.214e-06\\

    
    &CaraNet (MIIP'22) 
    ~\cite{lou2022caranet}
    &\textcolor{green}{0.6916}	&0.7642	&0.7639	&\textcolor{red}{0.8758}	&\textcolor{green}{0.9666} &0.7602 &3.224e-03\\ 

    &PVT-CASCADE (WACV'23) 
    ~\cite{rahman2023medical}
    &\textcolor{blue}{0.7288}	&\textcolor{blue}{0.8036}	&\textcolor{blue}{0.8244} &\textcolor{green}{0.8521}	&\textcolor{blue}{0.9705} &\textcolor{blue}{0.8102}
    &2.413e-01\\ 


   &Mamba-UNet (arXiv'24)     ~\cite{wang2024mambaUnet}
    &0.6769  &0.7545  &0.7586 	&0.8460	&0.9595 &0.7500 &1.406e-04\\

    &RMAMamba-T (arXiv'25)     ~\cite{zeng2025rmamamba}
    &0.6905 &\textcolor{green}{0.7651}	&\textcolor{green}{0.7851}	&0.8386	&0.9649 &\textcolor{green}{0.7683} &2.370e-03\\


    &\textbf{\ourmodel~(Ours)} & \textcolor{red}{0.7467} & \textcolor{red}{0.8209} & \textcolor{red}{0.8315} & \textcolor{blue}{0.8586}   & \textcolor{red}{0.9747} & \textcolor{red}{0.8241}
    &-\\

    \hline
            
    \multirow{8}{*}{\shortstack{PolypDB \\ (WLI)}} & 
    
    U-Net(MICCAI'15)
    ~\cite{ronneberger2015u}  
    & 0.8102 & 0.8762 & 0.8851 & 0.9096 & 0.9741 & 0.8765 &1.474e-12\\
    
    &U-Net++(TMI'19)
    ~\cite{zhou2018unetplus} 
    & 0.7784 & 0.8508 & 0.8649 & 0.8960   &0.9698 & 0.8537 
    &6.066e-17\\
    
    &PraNet (MICCAI'20)
    ~\cite{fan2020pranet} 
    & 0.8723 & 0.9246 & 0.9201 & \textcolor{green}{0.9441}  & \textcolor{green}{0.9847} & 0.9202 &6.109e-02\\

    
    &CaraNet (MIIP'22) 
    ~\cite{lou2022caranet}
    & 0.8747 & 0.9266 & 0.9281 & 0.9385  & 0.9841 & 0.9248
    &1.274e-01\\
     
    &PVT-CASCADE (WACV'23) 
    ~\cite{rahman2023medical}
    &\textcolor{red}{0.8927} & \textcolor{red}{0.9383} &\textcolor{blue}{0.9350} & \textcolor{red}{0.9503} &\textcolor{blue}{0.9858} & \textcolor{red}{0.9355}
    &3.382e-01\\


    &Mamba-UNet (arXiv'24)     ~\cite{wang2024mambaUnet}
    &0.8758  &0.9266  &0.9262 	&0.9422	&0.9838 &0.9243 &3.275e-02\\

    &RMAMamba-T (arXiv'25)     ~\cite{zeng2025rmamamba}
    &\textcolor{green}{0.8860}  &\textcolor{green}{0.9339}   &\textcolor{red}{0.9385} 	&0.9423	&\textcolor{red}{0.9860} &\textcolor{blue}{0.9351} &9.589e-01\\


    &\textbf{\ourmodel~(Ours)} & \textcolor{blue}{0.8862} & \textcolor{blue}{0.9342} & \textcolor{green}{0.9318} & \textcolor{blue}{0.9477}   & \textcolor{blue}{0.9858} & \textcolor{green}{0.9314}
    &-\\
  \hline
	\end{tabular}
\end{table*}

\begin{table*}[t!]
	\centering
 
	\caption{Quantitative results of different methods on the PolypDB dataset~\cite{tajbakhsh2015automated} with center-wise setting. 
	}\label{tab:center-wise}
	\begin{tabular}{clccccccc}
            \toprule
		
    Dataset & Methods 
    & mIoU ${\uparrow}$ 
    & mDSC ${\uparrow}$  
    & Recall ${\uparrow}$ 
    & Precision ${\uparrow}$ 
    & Accuracy ${\uparrow}$ 
    & F2 ${\uparrow}$
    & P-values ${\downarrow}$\\
    \hline

    \multirow{8}{*}{
    \shortstack{PolypDB \\ (BKAI)}} &
    
    U-Net (MICCAI'15)
    ~\cite{ronneberger2015u}  &0.7929	&0.8568	&0.9036	&0.8624	&0.9857 &0.8766
    &1.088e-26\\
    
    &U-Net++ (TMI'19)
    ~\cite{zhou2018unetplus} 	&0.7750	&0.8388	&0.8917	&0.8613	&0.9841 &0.8614
    &1.283e-33\\
    
    &PraNet (MICCAI'20)
    ~\cite{fan2020pranet} &0.8453	&0.9010	&\textcolor{green}{0.9271}	&0.9013	&0.9907 &0.9127
    &3.758e-05\\
    
    
    &CaraNet (MIIP'22) 
    ~\cite{lou2022caranet}
    &0.8458	&0.8998	&0.9207	&0.9083 &0.9911 &0.9091 
    &8.982e-05\\
    
    &PVT-CASCADE (WACV'23) 
    ~\cite{rahman2023medical}
    &\textcolor{red}{0.8640}	&\textcolor{red}{0.9138}	&0.9291
    &\textcolor{red}{0.9206} &\textcolor{red}{0.9924} &\textcolor{blue}{0.9199} 
    &6.100e-01\\ 


    &Mamba-UNet (arXiv'24)     ~\cite{wang2024mambaUnet}
    &0.8430  &0.8983  &\textcolor{green}{0.9301} 	&0.8951	&0.9905 &0.9117 &8.065e-07\\

    &RMAMamba-T (arXiv'25)     ~\cite{zeng2025rmamamba}
    &\textcolor{green}{0.8606}  &\textcolor{green}{0.9119}   &\textcolor{blue}{0.9302} 	&\textcolor{blue}{0.9151}	&\textcolor{green}{0.9920} &\textcolor{green}{0.9198} &7.491e-01\\

    &\textbf{\ourmodel~(Ours)} &\textcolor{blue}{0.8614}	&\textcolor{blue}{0.9127}	&\textcolor{red}{0.9314}	&\textcolor{green}{0.9137} &\textcolor{blue}{0.9921} &\textcolor{red}{0.9201} 
    &-\\ 

    \hline

    \multirow{8}{*}{
    \shortstack{PolypDB \\ (Karolinska)}} & 
    
    U-Net (MICCAI'15)
    ~\cite{ronneberger2015u}  
    &0.7499	&0.8303	&0.8521	&0.8880	&0.9681 &0.8369
    &1.603e-03\\
    
    &U-Net++ (TMI'19)
    ~\cite{zhou2018unetplus} 
    &0.6988	&0.7792	&0.8042	&0.8842	&0.9635 &0.7877
    &8.294e-04\\
    
    &PraNet (MICCAI'20)
    ~\cite{fan2020pranet} 
    &0.8827	&0.9304	&\textcolor{red}{0.9538}	&0.9243	&0.9879 &0.9400
    &4.610e-01\\

    
    &CaraNet (MIIP'22) 
    ~\cite{lou2022caranet}
    &0.8879	&0.9369	&0.9216	&0.9632 &0.9880 &0.9268
    &2.387e-01\\ 
    
    &PVT-CASCADE (WACV'23) 
    ~\cite{rahman2023medical}
    &\textcolor{red}{0.9152}	&\textcolor{red}{0.9547}	&\textcolor{blue}{0.9467}	&\textcolor{blue}{0.9665} &\textcolor{red}{0.9918} &\textcolor{red}{0.9495} 
    &3.573e-01\\ 
    
    
    &Mamba-UNet (arXiv'24)     ~\cite{wang2024mambaUnet}
    &0.8896  &0.9325  &0.9194 	&\textcolor{green}{0.9646}	&0.9870 &0.9238 &2.426e-01\\

    &RMAMamba-T (arXiv'25)     ~\cite{zeng2025rmamamba}
    &\textcolor{blue}{0.9116}  &\textcolor{blue}{0.9521}  &0.9405 	&\textcolor{red}{0.9682}	&\textcolor{blue}{0.9909} &\textcolor{blue}{0.9447} &4.421e-01\\
    	
    &\textbf{\ourmodel~(Ours)} &\textcolor{green}{0.9041}	&\textcolor{green}{0.9464}	&\textcolor{green}{0.9411}	&0.9601 &\textcolor{green}{0.9899} &\textcolor{green}{0.9425} 
    &-\\ 
    
    \hline

      \multirow{8}{*}{
      \shortstack{PolypDB \\ (Simula)}} & 
    
    U-Net (MICCAI'15)
    ~\cite{ronneberger2015u}  &0.7690	&0.8434	&0.8587	&0.8923	&0.9679 &0.8463
    &3.119e-10\\
    
    &U-Net++ (TMI'19)
    ~\cite{zhou2018unetplus} 	&0.7647	&0.8374	&0.8566	&0.8842	&0.9674 &0.8426 &4.510e-10\\

    &PraNet (MICCAI'20)
    ~\cite{fan2020pranet} &0.8582	&0.9166	&0.9172	&0.9347	&0.9815 &0.9145
    &3.245e-01\\


    &CaraNet (MIIP'22)
    ~\cite{lou2022caranet}
    &0.8540	&0.9111	&0.9122	&0.9348 &0.9808 &0.9086 
    &1.271e-01\\ 
    
    &PVT-CASCADE (WACV'23) 
    ~\cite{rahman2023medical}
    &\textcolor{red}{0.8795}	&\textcolor{red}{0.9306}	&\textcolor{red}{0.9316}	&\textcolor{blue}{0.9424} &\textcolor{red}{0.9854} &\textcolor{red}{0.9295} 
    &2.135e-01\\  


    &Mamba-UNet (arXiv'24)     ~\cite{wang2024mambaUnet}
    &0.8580  &0.9151  &\textcolor{green}{0.9224} 	&0.9288	&0.9801 &\textcolor{green}{0.9169} &2.789e-01\\

    &RMAMamba-T (arXiv'25)     ~\cite{zeng2025rmamamba}
    &\textcolor{green}{0.8698}  &\textcolor{green}{0.9202}  &0.9173 	&\textcolor{red}{0.9474}	&\textcolor{blue}{0.9820} &0.9167 &6.892e-01\\

    &\textbf{\ourmodel~(Ours)} &\textcolor{blue}{0.8712}	&\textcolor{blue}{0.9230}	&\textcolor{blue}{0.9239}	&\textcolor{green}{0.9406} &\textcolor{green}{0.9808}
    &\textcolor{blue}{0.9216} 
    &-\\ 
    
    \hline

	\end{tabular}
\end{table*}

\subsection{Quantitative benchmarking}
Here, we compare the proposed approach with SOTA models across three major architectural categories:  CNN based methods (U-Net~\cite{ronneberger2015u}, U-Net++~\cite{zhou2018unetplus}, PraNet~\cite{fan2020pranet}, 
and CaraNet~\cite{lou2022caranet}) Transformer based approaches (PVT-CASCADE~\cite{rahman2023medical}) and recently popular mamba based approaches (Mamba-UNet~\cite{wang2024mambaUnet} and RMAMamba-T~\cite{zeng2025rmamamba}) with the modality-wise and the center-wise setting. In the center-wise setting, the images are categorized based on their corresponding centers. Since only WLI modality is included in all centers, the results can reflect the model's learning ability and highlight differences caused by the devices used across different centers. In the modality-wise setting, polyp images are categorized based on their respective modalities, which enables the evaluation of the model's performance across different modalities.


\begin{figure}[t!]
	\centering
	\includegraphics[width=1.0\linewidth]{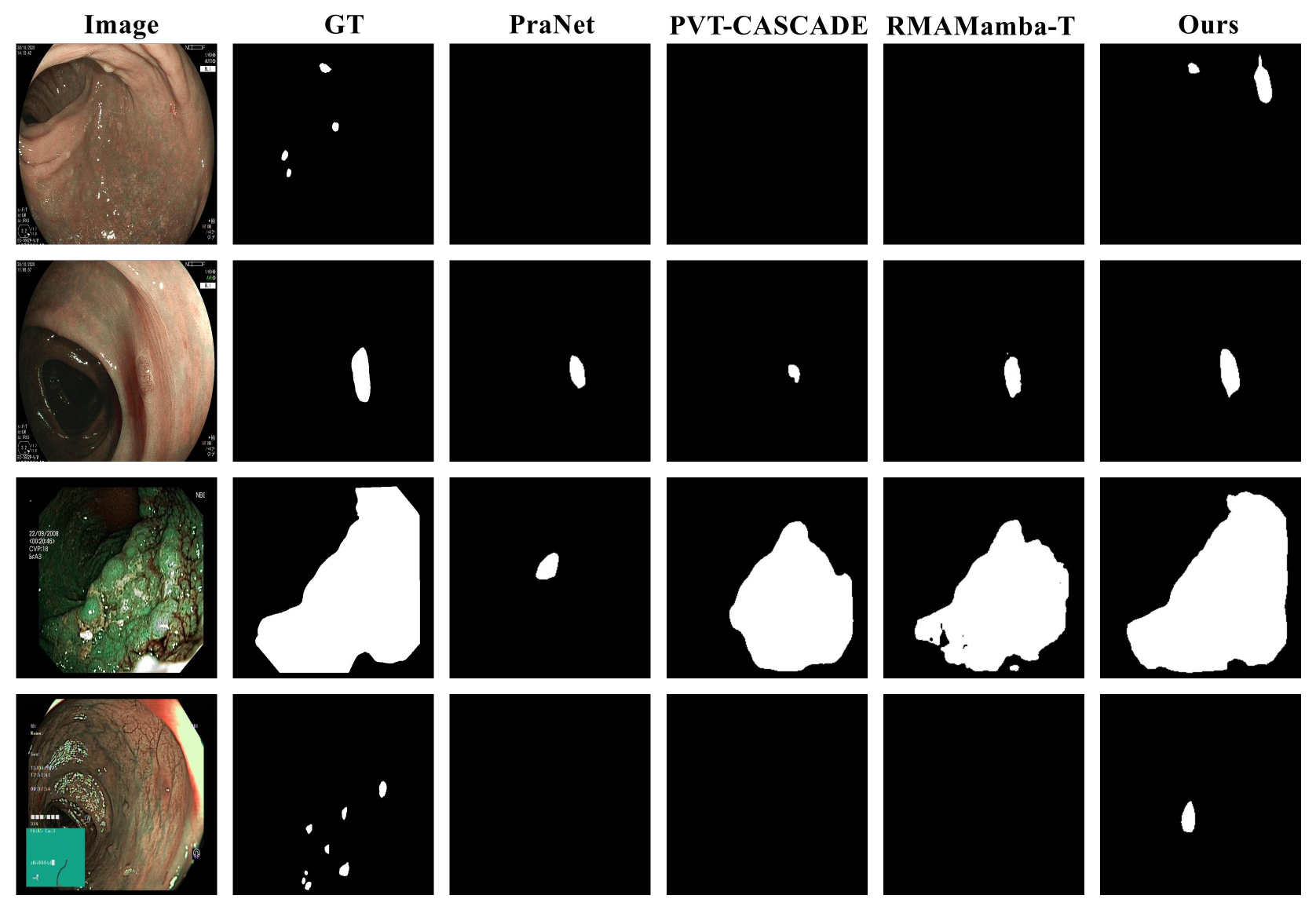}
	\caption{Qualitative results of different methods across BLI and NBI modalities. It is obvious that FocusNet produces better segmentation masks, particularly in challenging scenarios involving small, flat, multiple polyps. Models such as RMAMamba-T showed under-segmentation whereas our model provided accurate and robust segmentation.}
    
    \label{fig:Results_BLI_NBI}
\end{figure}

\begin{figure}[t!]
	\centering
\includegraphics[width=1.0\linewidth]{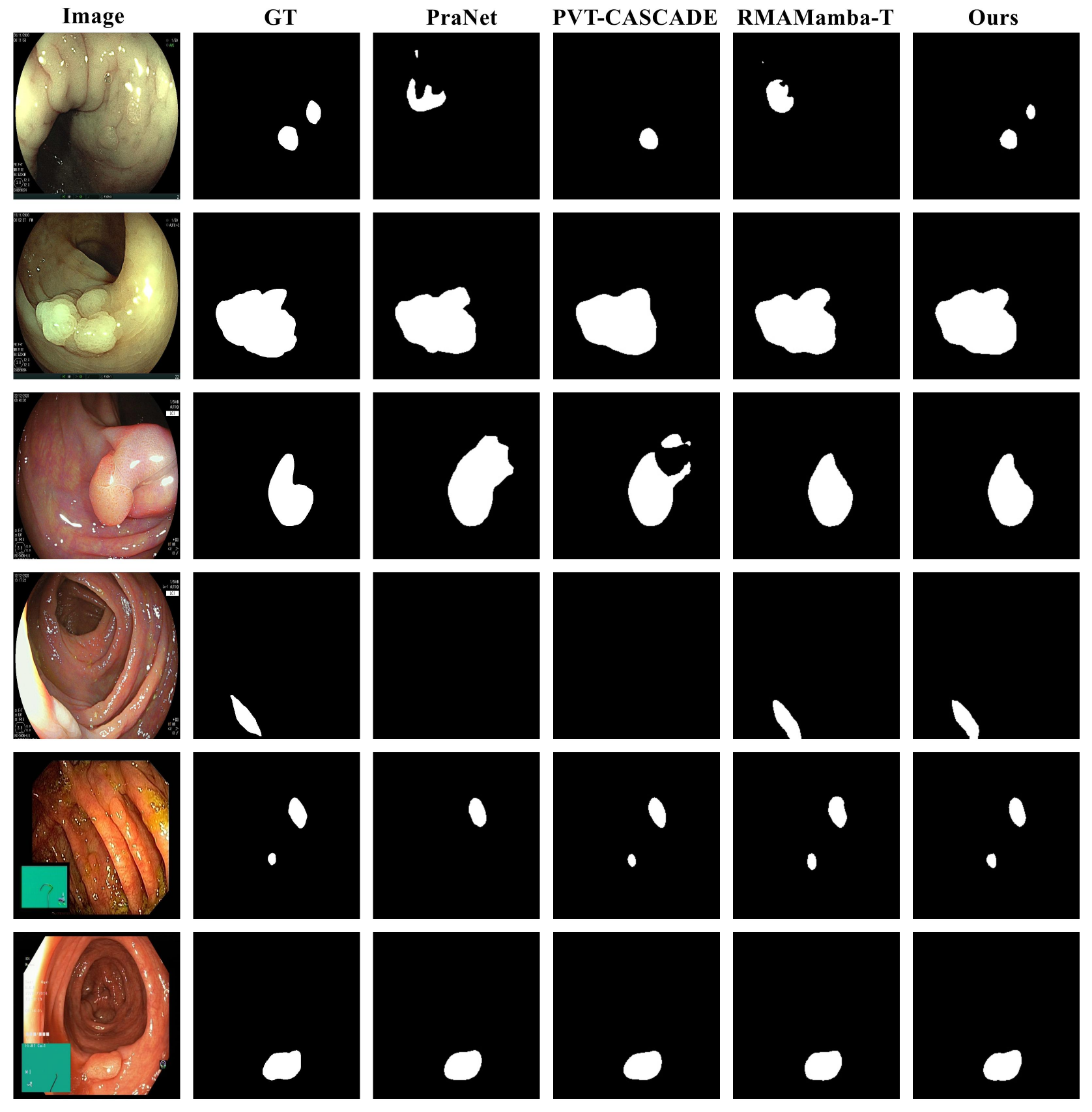}
	\caption{Qualitative results of different methods across FICE, LCI, and WLI modalities. Here, most of the advanced models (PVT-Cascade and RMAMamba-T) can detect polyps, however, FocusNet produces segmentation masks with higher boundary precision and consistency.}

        \label{fig:Results_FICE_etc}
\end{figure}
\subsubsection{Quantitative Comparison}
Tables~\ref{tab:modality-wise} and \ref{tab:center-wise} present the quantitative results of \ourmodel~and seven SOTA methods on the PolypDB dataset with modality-wise and center-wise settings. In the modality-wise setting, it is evident that our \ourmodel~outperforms other cutting-edge methods on modalities such as BLI, FICE, LCI, and NBI. For instance, the classic method PraNet obtains 0.6582, 0.7353, 0.7660, and 0.8430 in mIoU, mDSC, Recall, and Precision on BLI dataset, and 0.6360, 0.7131, 0.7332, and 0.8419 on NBI dataset, respectively. Our \ourmodel~shows a significant improvement over PraNet. In the BLI and NBI datasets, our model outperforms PraNet by 8.94\% and 10.78\% in mDSC, 8.78\% and 11.07\% in mIoU, and 8.12\% and 9.83\% in Recall. In addition, our \ourmodel~outperforms the most advanced method, RMAMamba-T, across all evaluation metrics. For instance, on the BLI dataset, FocusNet surpasses RMAMamba-T by 4.22\%, 4.25\%, and 3.49\% in mIoU, mDSC, and Recall, respectively. On the NBI dataset, our \ourmodel~outperforms G-CASCADE by approximately 5.62\%, 5.58\%, and 4.64\% in mIoU, mDSC, and Recall, respectively.

In the WLI dataset, \ourmodel's performance is slightly inferior to the advanced method PVT-CASCADE, falling short by 0.41\% and 0.65\% in mDSC and mIoU, respectively. In the center-wise experiment, \ourmodel~'s performance is slightly lower than that of PVT-CASCADE. In the BKAI center, it falls short by 0.11\% and 0.26\% in mDSC and mIoU, respectively. In the Simula center, FocusNet falls short by 0.67\% and 0.83\% in mDSC and mIoU, respectively. This indicates that \ourmodel~demonstrates strong generalization, but its learning ability has not exhibited the best performance among SOTA models. A potential reason for this phenomenon is that although the combination of local and pooling attention enhances the model's generalization, along with shallow information supplementing boundary details, pooling attention can also lead to the loss of other fine-grained details, ultimately reducing the model's learning ability. 



\subsubsection{Qualitative Comparison}
\begin{figure}[t!]
	\centering
	\includegraphics[width=\linewidth]{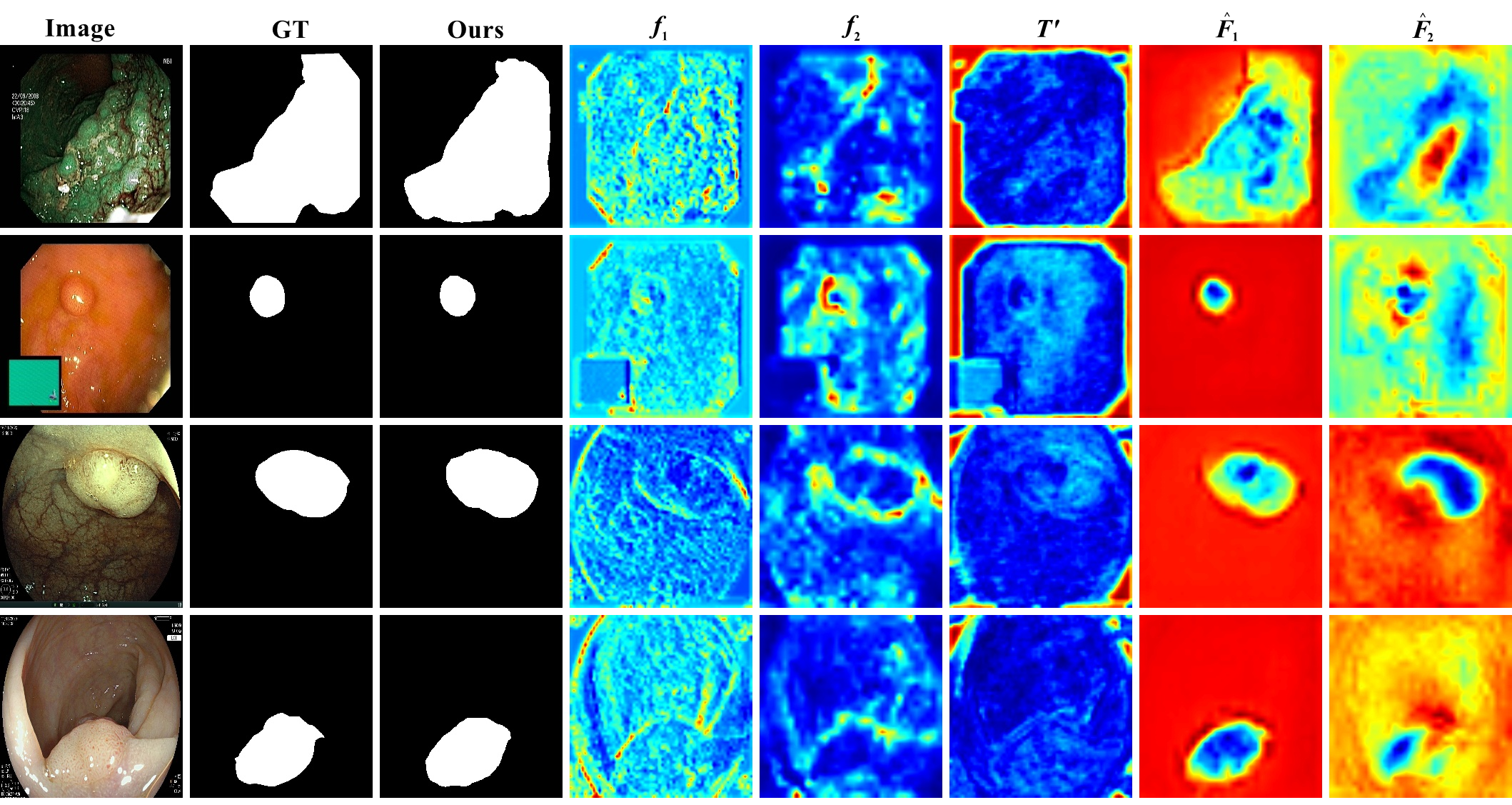}
	\caption{Visualization of intermediate feature maps from our model. Columns from left to right: input image, ground truth mask (GT), predicted mask (Ours), encoder feature maps ($f_1$, $f_2$), detail-enhanced transformer map ($T'$), and decoder outputs ($\hat{F}_1$, $\hat{F}_2$). Here, the shallow feature $f_1$ preserves fine-grained details, while $f_2$ captures higher-level semantic context. The enhanced map $T'$ refines boundary structures, and the decoder outputs integrate local and global cues for precise segmentation. The progression illustrates how FocusNet effectively combines detail, context, and attention to delineate polyps.} 
    \label{fig:heatmap}
\end{figure}

To ensure a comprehensive and intuitive comparison, we present the predictions of different models along with their corresponding images and ground truths from different datasets. Figure~\ref{fig:Results_BLI_NBI} demonstrates the segmentation results of challenging images from the BLI and NBI datasets. Our method effectively distinguishes complex scenes containing multiple tiny polyps within a single image as shown in the first and fourth rows), whereas most methods fail to detect the polyp regions. The results in the second row indicate that~\ourmodel~exhibits a broader detection range while successfully identifying highly camouflaged polyps. In the third row, we observe that mixed blood stains and unknown secretions in the polyp images present significant challenges for segmentation. While most models struggle to accurately differentiate lesion areas, our method successfully detects the majority of polyp regions, achieving results closely aligned with expert annotations. Meanwhile, in Figure~\ref{fig:Results_FICE_etc} , the images depict relatively conventional and typical cases, where most models can detect polyp regions effectively. In this scenario, our model demonstrates minimal deviation from expert annotations across all samples and significantly outperforms other methods. To better describe the feature propagation, we present the visualization of intermediate feature maps from our model in Figure ~\ref{fig:heatmap}.




\subsection{Ablation Study}
We conduct extensive ablation experiments to evaluate the effectiveness of each of the three proposed modules in FocusNet. We employ the same training strategy in the ablation experiments to ensure fairness while evaluating on PolypDB modality-wise setting. Starting with PVTv2 as the backbone of \ourmodel~and CIDM as the base decoder, we progressively incorporate the DEM and FAM into the backbone. This results in six variants of the model: backbone + CIDM-M, backbone + CIDM-A, backbone + CIDM-M + FAM, backbone + CIDM-A + FAM, backbone + CIDM-M + CIDM-A + FAM and backbone + CIDM-M + CIDM-A + FAM + DEM. As shown in Table~\ref{tab:ablation}, the integration of three modules enables \ourmodel~to achieve the best performance demonstrating the effectiveness of each component in the network. 

Table~\ref{tab:complexity} presents the computational complexity analysis for each model. While FocusNet achieves higher performance, it also has a larger number of parameters and computational costs. This trade-off between model accuracy and efficiency can be justified by considering performance gains. 

\begin{table}[t!]
	\centering
		\caption{Computational complexity of various Methods.}
    \label{tab:complexity}
    \resizebox{\columnwidth}{!}{
	\begin{tabular}{l|c|c|c}
        \toprule	
		Method 
		& Parameters 
		& Flops (GMac) 
		& FPS  \\
		
		\hline
		
		UNet~\cite{ronneberger2015u}
		&31.04M
		&48.23      
            &259.47\\
		
		U-Net++~\cite{zhou2018unetplus}
		&47.18M
		&135.19      
            &168.37\\
		
		PraNet~\cite{fan2020pranet}
		&32.55M
		&6.92     
            &48.08\\
		
		UACANet~\cite{kim2021uacanet}
		&69.16M
		&31.58          &28.52\\
				
		CaraNet~\cite{lou2022caranet}
		&46.64M
		&11.45      
            &27.02\\

		PVT-CASCADE~\cite{rahman2023medical}
		&35.27M
		&8.39           &45.61\\

            Mamba-UNet~\cite{wang2024mambaUnet}
		&19.12M
		&5.87           &54.83\\

		RMAMamba-T~\cite{zeng2025rmamamba}
		&30.13M
		&6.49           &38.24\\
        
        
		\textbf{\ourmodel~        (Ours)}
		&63.41M
		&14.77      
            &18.58 \\
        \bottomrule
	\end{tabular}
       }
\end{table}

\begin{table}[t!]
	\centering
	\caption{Ablation study on the PolypDB showing the impact of different modules.}
    \label{tab:ablation}
    \resizebox{\columnwidth}{!}{
	\begin{tabular}{c|l|cc}
        \toprule	
		\multirow{2}{*}{Exp.} 
		& \multirow{2}{*}{Setting} 
		& \multicolumn{2}{c}{PolypDB (BLI)} \\
		& & mIoU ${\uparrow}$ & DSC ${\uparrow}$ \\
		
		\hline
		
		\#1
		& Backbone + CIDM-M
		&0.7373 &0.8170 \\
		
		\#2
		& Backbone + CIDM-A
		&0.7232 &0.8100 \\
		
		\#3
		& Backbone + CIDM-M + FAM 
		&0.7418 &0.8214 \\
		
		\#4
		& Backbone + CIDM-A + FAM
		&0.7332 &0.8133  \\

		\#5
		& Backbone + CIDM-M + CIDM-A + FAM
		&0.7356 &0.8141 \\
		
		\textbf{Ours}
		& Backbone + CIDM-M + CIDM-A + FAM + DEM
		&\textcolor{red}{0.7460} &\textcolor{red}{0.8247} \\
        \bottomrule
	\end{tabular}
       }
\end{table}

\begin{figure}[t!]
	\centering
	\includegraphics[width=0.8\linewidth, trim=0cm 10mm 0cm 0cm, clip]{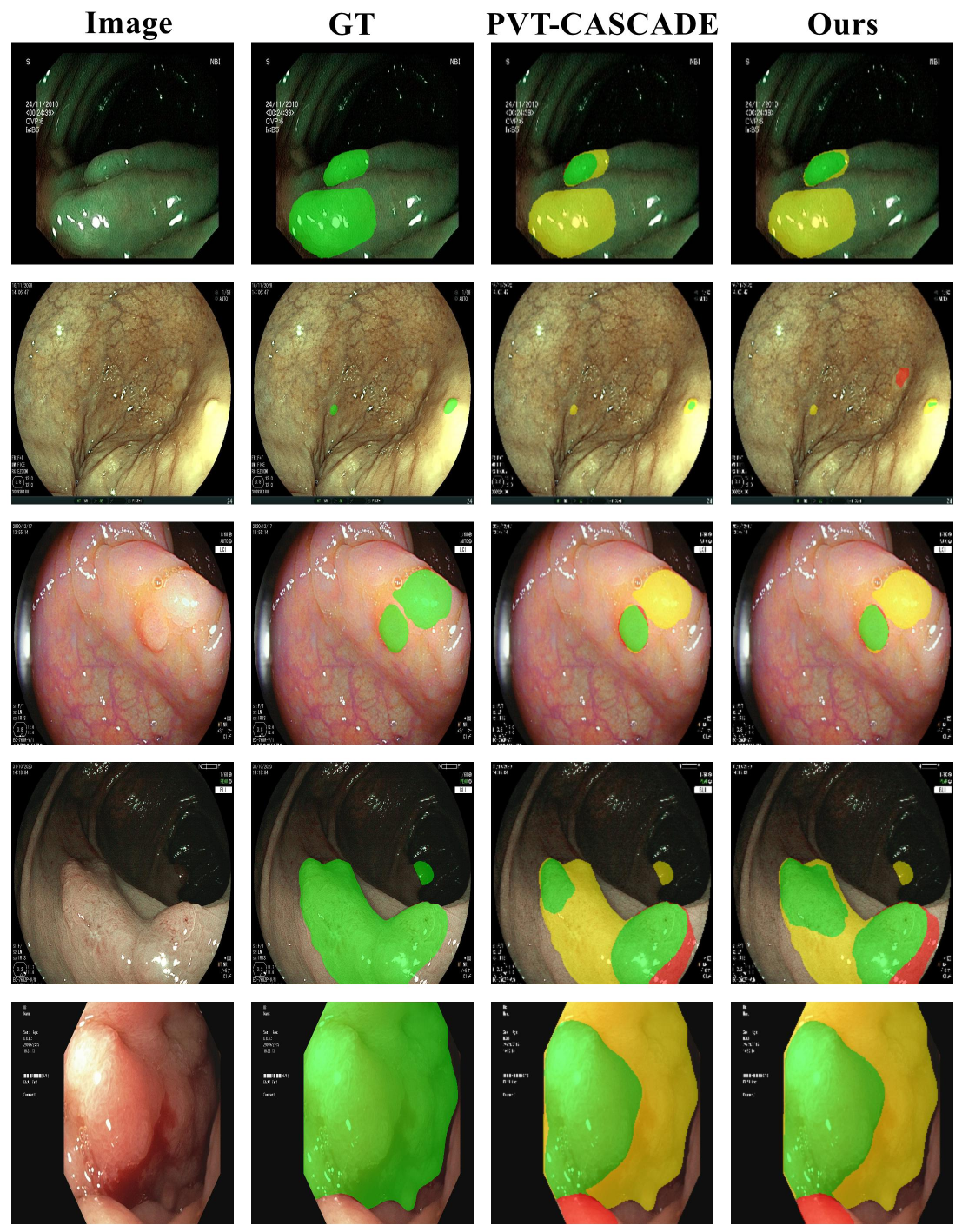}
	\caption{Qualitative visualization of some failure cases.} 
    \vspace{-4mm}
	\label{fig:Results_hard}
\end{figure}

\section{Discussion}\label{disscussion}
The model proposed combines local and pooling attention mechanisms to enhance feature representation by integrating boundary information from shallow features and has achieved superior segmentation performance. Both the local and pooling window sizes can be adjusted to optimize the model for specific tasks. The model achieves SOTA performance across data from different modalities and centers on the PolypDB dataset, validating the effectiveness of the FocusNet.

From Table~\ref{tab:modality-wise}, it can be observed that FocusNet is statistically significant across most modalities (for example, BLI, NBI, etc.), with low P-values ($P < 0.05$) when compared to SOTA models. However, our model may still perform poorly in certain scenarios. Figure~\ref{fig:Results_hard} presents several failure cases, where green indicates the correct polyp, yellow represents the missed polyp, and red indicates the incorrect polyp. In the first and third rows, polyps affected by reflection or mucus were not successfully detected, which could lead to missed diagnoses in clinical settings. In the fourth and fifth rows, while both PVT-CASCADE and \ourmodel~exhibit large areas of missed polyps, \ourmodel~presents a larger correctly predicted area and smaller missed polyp areas compared to PVT-CASCADE. 

Additionally, in the fifth-row results, \ourmodel~also provides a more complete and accurate prediction of the left boundary. Notably, in the fourth-row results, neither model detects polyps in the low-light areas, further highlighting how lighting can significantly impact model performance. It is worth noting that in the second row, while the red area predicted by FocusNet is not actual polyp tissue, its unusual appearance and structure make it highly suspicious and clinically valuable. This demonstrates FocusNet’s high sensitivity to potentially abnormal tissues. These failure cases highlight both the limitations and advantages of our model. In the future, we aim to design and optimize models to address challenges related to missed and misdiagnosed cases caused by factors such as lighting, mucus, and secretions.

From the quantitative and qualitative analysis, we observed that FocusNet can robustly segment polyps across multiple imaging modalities (WLI, BLI, NBI, LCI, and FICE) and varied clinical centers can have direct implications for CRC screening and prevention. By reducing the risk of missed polyps—especially small, flat, or camouflaged lesions—FocusNet can act as a foundation for the future development of real-time, multi-modal computer-aided diagnosis systems. 

\section{Conclusion}\label{conclusion}
We proposed FocusNet for accurate polyp segmentation. The cross-semantic interaction decoder module was designed to enhance the network's scene perception ability by promoting more effective information interaction across both the channel and spatial dimensions between the features. Additionally, the detail enhancement module was proposed to refine detail information through multi-path deformable convolutions. Furthermore, the network utilized a novel focus attention module that combined local and pooling attention mechanisms by adjusting their respective window sizes to balance the local details and global context. Extensive experiments on PolypDB demonstrated that FocusNet significantly outperformed current SOTA models. Although FocusNet demonstrated excellent segmentation performance, there were still some challenging cases where the model struggled to identify and predict accurate polyp boundaries. In the future, we will test our architecture with video sequence datasets for temporal consistency. 

\bibliographystyle{IEEEtran}
\bibliography{references}{}

\end{document}